\documentclass[12pt,preprint]{aastex}

\shorttitle{EUV tornadoes}

\shortauthors{Jun Zhang and Yang Liu}

\begin{document}

\title{Ubiquitous rotating network magnetic fields and EUV cyclones
in the quiet Sun}

\author{Jun Zhang\altaffilmark{1} and Yang Liu\altaffilmark{2}}

\altaffiltext{1}{Key Laboratory of Solar Activity, National
Astronomical Observatories, Chinese Academy of Sciences, Beijing
100012, China; E-mail: zjun@nao.cas.cn}

\altaffiltext{2}{W. W. Hansen Experimental Physical Laboratory,
Stanford University, Stanford, CA 94305, USA; E-mail:
yliu@sun.stanford.edu}

\begin{abstract}

We present the {\it Solar Dynamics Observatory} (SDO) Atmospheric
Imaging Assembly (AIA) observations of EUV cyclones in the quiet
Sun. These cyclones are rooted in the Rotating Network magnetic
Fields (RNFs). Such cyclones can last several to more than ten
hours, and, at the later phase, they are found to be associated with
EUV brightenings (microflares) and even EUV waves. SDO Helioseismic
and Magnetic Imager (HMI) observations show an ubiquitous presence
of the RNFs. Using HMI line-of-sight magnetograms on 2010 July 8, we
find 388 RNFs in an area of 800$\times$980 square arcseconds near
the disk center where no active region is present. The sense of
rotation shows a weak hemisphere preference. The unsigned magnetic
flux of the RNFs is about 4.0$\times$10$^{21}$ Mx, or 78\% of the
total network flux. These observational phenomena at small scale
reported in this letter are consistent with those at large scale in
active regions. The ubiquitous RNFs and EUV cyclones over the quiet
Sun may suggest an effective way to heat the corona.

\end{abstract}


\section{INTRODUCTION}

In the past 20 years, observations from space-based telescopes
revealed that the solar corona and transition region are much more
dynamic than had been thought. Surges, jets, macrospicules, bright
points and macroflares take place everywhere in the solar
atmosphere. Sometimes, surges, jets and macrospicules exhibit
rotating motions \citep{pik98, zha00, pat08, 2009ApJ...707L..37L,
2010A&A...510L...1K}, which is suggested to be an indicator of the
existence of Alfv\'{e}n waves in these structures (Sterling, 1998).
The mechanism providing such a highly dynamic corona and transition
region is considered to be related to the long-standing puzzle of
how the corona is heated to a million Kelvin \citep{asc04}.

Rotational motions in the photosphere (Brandt et al 1988) may create
rotating motions of surges (jets) and macrospicules which are driven
by localized downdrafts that collect the cold plasma returning to
the solar interior after releasing internal energy (e.g., Spruit et
al. 1990; Stein \& Nordlund 1998). Usually those vortices are small
(0.5 Mm) and last several minutes, but sometimes they could be large
and survive much longer (see, e.g. Attie et al. 2009). In the quiet
Sun, about 3.1$\times$10$^{-3}$ vortices Mm$^{-2}$  minute$^{-1}$
are found \citep{bon10}, and they do not have a preferred rotation
sense on the different hemispheres (Bonet et al. 2008). The vortices
on the photosphere can propagate upward along the field lines (e.g.,
Choudhuri et al. 1993; Zirker 1993; van Ballegooijen et al. 1998),
and govern the evolution of magnetic footpoints \citep{bal10}. In
the low chromosphere, vortical motions with possible propagation of
waves were also detected for the first time by Wedemeyer-B{\"o}hm \&
Rouppe van der Voort (2009). Moreover, they often wind up opposite
polarity field lines, facilitating magnetic reconnection and ensuing
energy release.  For active regions, rotating motion is one of the
most important processes for sunspot evolution, and may be
responsible for the flaring \citep{ger03}. Thus those rotating
motions may provide a mechanism to build up the magnetic energy in
the corona that is eventually released during transients, which in
turn heats the corona.

In this Letter, we report the discovery of EUV cyclones and Rotating
Network magnetic Fields (RNFs) in the quiet Sun with the
observations from the Atmospheric Imaging Assembly (AIA; Lemen et
al. 2011) and the Helioseismic and Magnetic Imager (HMI; Wachter et
al. 2011) aboard the {\it Solar Dynamics Observatory} (SDO). The
paper is organized as follow. In Section 2 we describe the
observational data we use. An analysis of the data is presented in
Section 3. In Section 4, we conclude this study and discuss the
results.

\section{OBSERVATIONS}

SDO/AIA observes uninterruptedly the Sun's full disk at 10
wavelengths at a 12-second cadence and a 0.6$''$ pixel$^{-1}$
sampling. The measurements reflect various temperatures of the solar
atmosphere (from $\sim$5000 K to $\sim$2.5 MK) from the photosphere
to the corona. SDO/HMI, from polarization measurements, information
such as Doppler-velocity, line-of-sight (LOS) magnetic field
strength, and vectorial information of the magnetic field are
obtained. The data cover the full disk of the Sun with a spatial
sampling of 0.5 $''$ pixel$^{-1}$. The full disk LOS magnetograms,
used in this Letter, are taken at a cadence of 45 seconds.

\section{ANALYSIS}

The phenomenon of the EUV cyclones is seen everywhere in the quiet
Sun in the AIA data in all EUV channels of 171, 193, 304, 211, 131,
335, and 94 {\AA}. Those cyclones are found to be associated with
the magnetic fields that show conspicuous rotation when viewed with
the movies of the HMI time-series magnetograms. We use the term RNFs
to refer to those rotating magnetic fields. Here we present two
examples of the EUV cyclones to demonstrate their evolutionary
characteristics and properties. Shown in Figures. 1(A) and 1(B) is a
cyclone occurred on July 2 (see also movie1). The images were taken
in the 171 {\AA} channel. This cyclone was in the northern
hemisphere, close to the solar equator. It started at $\sim$2:00 UT,
rotated counter-clockwisely, and lasted 12 hours. The footpoint of
the cyclone on the photosphere, shown as a positive field patch in
the HMI LOS magnetograms (the white feature in Figures. 1(C) and
1(D)), also rotated counter-clockwisely. At 04:02:48 UT, the
magnetic flux in this patch was 2.9$\times$10$^{19}$ Mx, one order
greater than the flux in a typical network element($\sim$10$^{18}$
Mx, Zhang et al. 2006). The noise level of HMI 45 s magnetograms is
10.2 Mx cm$^{-2}$ (Liu et al. 2011), and the minimum flux obtained
for this level in HMI is 1.3 $\times$ 10$^{16}$ Mx. Its longer axis
was along the northeast-southwest direction. At 06:44:48 UT, this
axis rotated to southeast-northwest direction. It suggests that the
magnetic field rotated 64$^{\circ}$ within 162 minutes.

The other example was in the southern hemisphere on July 20. It
occurred at 07:00 UT, and lasted 9 hours. Figures. 2 (A) and (B)
show the cyclone at the 171 {\AA} images. The two circles denote the
place where a time-slice map is made. Different from the cyclone on
July 2, this cyclone rotated clockwisely (see movie2). From 11:23:48
UT to 13:20:48 UT, the positive magnetic fields (black contours in
Figures 2(C) and 2(D)), which the cyclone are rooted in, rotated
83$^{\circ}$ clockwisely (see (C) and (D) in Figure 2) with a speed
of 0.8$^{\circ}$ min$^{-1}$. Its flux is about 1.6$\times$10$^{19}$
Mx and then it cancelled with the nearby negative elements (white
contours in Figures 2(C) and 2(D)). We find that the negative
polarity with an absolute magnetic flux of 3.6$\times$10$^{18}$ Mx
disappeared and the positive polarity faded about
3.9$\times$10$^{18}$ Mx during the cancellation. The blue curve in
panel (D) is the contour of EUV brightening in panel (B). We can see
that the brightening corresponds to the neutral region, instead to
the positive polarity. Figure 2 (E) shows a time-slice map of the
AIA 171 {\AA} images. The X-axis refers to time, running from 2010
July 20 11:00 UT to 15:00 UT. The Y-axis refers to the angle
subtended by the dotted curves measured clockwise, whose origin is
the center of the circle and the reference direction is the West, as
shown in Panel (A). The dashed curve outlines the intersection of
the leading boundary of the cyclone. In the 3-hour time interval
(from 11:00 UT to 14:00 UT), the cyclone rotated about
360$^{\circ}$. In the later phase of the cyclone at 14:26:59 UT, an
EUV brightening appeared. This brightening continuously developed,
exhibiting subtle structures, and becoming a two-ribbon microflare
(the white patches in Figure 3). It lasted 2 hours, and disappeared
at 16:18:11 UT. Meanwhile a small-scale wave was triggered by the
microflare. It propagated with an average speed of 45 km s$^{-1}$,
and lasted almost 8 minutes. Its disturbance spread over an area of
6.3$\times$10$^{8}$ km$^{2}$, equivalent to a typical supergranular
cell.

Figure 4(A) shows the angular speed of the cyclone measured from the
time-slice map. It indicates that the cyclone underwent two
acceleration processes. At first the cyclone rotated with an angular
speed of 1$^{\circ}$ min$^{-1}$, 50 minutes later the speed reached
7$^{\circ}$ min$^{-1}$, and then slowed down to 1$^{\circ}$
min$^{-1}$ again. At 13:20 UT, the cyclone accelerated again. The
speed reached another peak (4.8$^{\circ}$ min$^{-1}$). Figure 4(B)
shows the propagating speed of the EUV wave. We determine the wave
front (denoted by stars in each panel in Figure 3) every 36-second,
and then compute the mean speed in the 36-second period. The
propagation speed of the wave was not constant. There were two
peaks, ($\sim$80 km s$^{-1}$ at 16:00:06 UT, and 85 km s$^{-1}$ at
16:05:30 UT).

Observations have shown that EUV cyclones are rooted in the RNFs. An
interesting question is then: how many RNFs are there on the solar
surface? To answer this question, we surveyed the 2010 July 8
AIA/HMI data when no active region was present on the solar disk. In
an area of 800$\times$980 square arcseconds near the disk center, we
found 388 RNFs (movie3 shows an example of the RNFs). The occurrence
of these events is 4.9 $\times$ 10$^{-4}$ arcsec$^{-2}$ day$^{-1}$.
The average lifetime of the RNFs is about 3.5 h. The distribution of
these RNFs is plotted in Figure 5. Their magnetic fluxes range from
1.24$\times$10$^{17}$ Mx to 5.85$\times$10$^{19}$ Mx. The average
flux is 1.03$\times$10$^{19}$ Mx, and the total flux of these RNFs
is about 4.0$\times$10$^{21}$ Mx, or 78\% of the unsigned magnetic
flux in that area. There are 179 and 209 RNFs in the northern and
southern hemispheres, respectively. The sense of the rotation shows
a weak hemisphere preference: 61\% of the RNFs in the northern
hemisphere (109 of the 179 RNFs) rotated counter-clockwisely, while
56\% of the RNFs in the northern hemisphere (116 of 209 RNFs)
rotated clockwisely. It leads to 5600 RNFs over the entire solar
surface each day. It infers that a total flux of
5.8$\times$10$^{22}$ Mx is rotating. Simultaneous AIA observations
show that among the cyclones relevant to these RNFs, 231 ones
($\sim$60\% of the total) are associated with microflares.

\section{DISCUSSION AND CONCLUSIONS}

In this Letter we report the discovery of EUV cyclones in the quiet
Sun. They are rooted in the RNFs, and last several to more than ten
hours, evidently different from surges (jets) and macrospicules,
which have lifetimes less than one hour \citep{zhan00}. About 60\%
EUV cyclones are associated with microflares and even small-scale
EUV waves at the later phase of cyclones. The microflares correspond
to the neutral region, instead to one polarity. HMI observations
show the ubiquitous presence of the RNFs. The sense of rotation
shows a weak hemisphere preference, which is also presented with
G-band observations in Vargas Dom{\'{\i}}nguez et al. (2011). The
average magnetic flux is 1.03$\times$10$^{19}$ Mx. Every day, we
infer that about 5600 RNFs are present over the entire solar
surface. The total unsigned flux is about 5.8$\times$10$^{22}$ Mx.

As revealed in this study, the brightenings (microflares) are not
corresponding to one polarity but to the neutral place where the
opposite polarities cancelled. This is consistent with the
morphological model shown in Figure 8 of Zhang et al. (2000b)., i.e,
when two network elements with opposite polarities converge together
and begin to cancel each other, a microflare appears at the
cancelling site.

Surface motion on the photosphere is one of the two mechanisms
proposed to build up the free energy in the corona (the other is the
emergence of magnetic field). It can produce current sheets parallel
to a separatrix that detach the interacting magnetic flux regions
(Sweet 1969; Lau 1993; see Somov 2006 for a review). In the
simulation of Gerrard et al. (2003), rotation of a sunspot with
inflow of a pore leads to a strong build-up of current which is
needed for magnetic reconnection. This current buildup is quite
similar to the conclusion of \cite{zha07, zha08}, i.e., the
rotational motions of sunspots relate to the transport of magnetic
energy and complexity from the low atmosphere to the corona and play
a key role in the onset of flares. In the quiet Sun, the random walk
of the footpoints of coronal loops in the solar granulation is
expected to cause the braiding of the field, which in turn leads to
a multitude of coronal reconnection events (Schrijver 2007).
Magnetic reconnection between sheared magnetic loops, indicating the
injection of magnetic helicity of mixed signs, works as a trigger
mechanism of solar flares \citep{kus04,kus05}. Thus, the phenomena
of RNFs, as well as the EUV cyclones, reported in this Letter,
likely contain the process of energy buildup and release in solar
corona.

What heats the solar corona remains one of the most important
questions in solar physics and astrophysics. \citet{kli06} has
proposed that three parts are involved to solve the coronal heating
puzzle, which are (1) identifying a source of energy and a mechanism
for converting the magnetic energy into heat, (2) determining how
the plasma responds to the heating, and (3) predicting the spectrum
of emitted radiation and its manifestation as observable quantities.
Our results may provide evidence to address the first two parts.
Continuous development of RNFs braids field lines of the magnetic
elements, manifested by the cyclones seen in the EUV observation.
Magnetic reconnection then takes place in the braiding fields,
releasing the stored energy that heats the corona. It is supported
by the observed brightenings (microflares) and EUV waves.

Up to now, one of the most common mechanisms to heat the corona is
the impulsive energy release (nanoflares, with times of minutes, and
dimensions of Mms) that occurs in the small-scale magnetic fields
(Parker 1988). AIA data indicate that microflares (or nanoflares)
are not impulsive events. Furthermore, they are accompanied with
other activities. For example, in the 2010 July 20 cyclone, the
microflare occurred 7 hours after the cyclone, followed by a
small-scale EUV wave. Small-scale waves in the quiet Sun may play a
role in coronal heating: they transport energy to other places.

The small-scale EUV waves reported here propogate at a speed of
35--85 km s$^{-1}$, much slower than the EUV waves originate from
active regions, which is in arange of 200-400 km s$^{-1}$ (see,
e.g., Thompson \& Myers 2009). The nature of the small-scale waves
is an interesting question that needs further investigation. It may
provide additional information that helps understand the EUV waves
that is still under intense debate (Wills-Davey \& Thompson 1999; Wu
et al. 2001; Ofman \& Thompson 2002; Schmidt \& Ofman 2010;
Delann\'{e}e 2000; Attrill et al. 2007; Attrill 2010; Liu et al.
2010).

\acknowledgments

The authors are indebted to the {\it SDO} teams for providing the
data. This work is supported by the National Natural Science
Foundations of China(G11025315, 40890161 and 10921303), the CAS
Project KJCX2-YW-T04, and the National Basic Research Program of
China under grant G2011CB811403.

\clearpage

\begin{figure}
\epsscale{1.0} \plotone{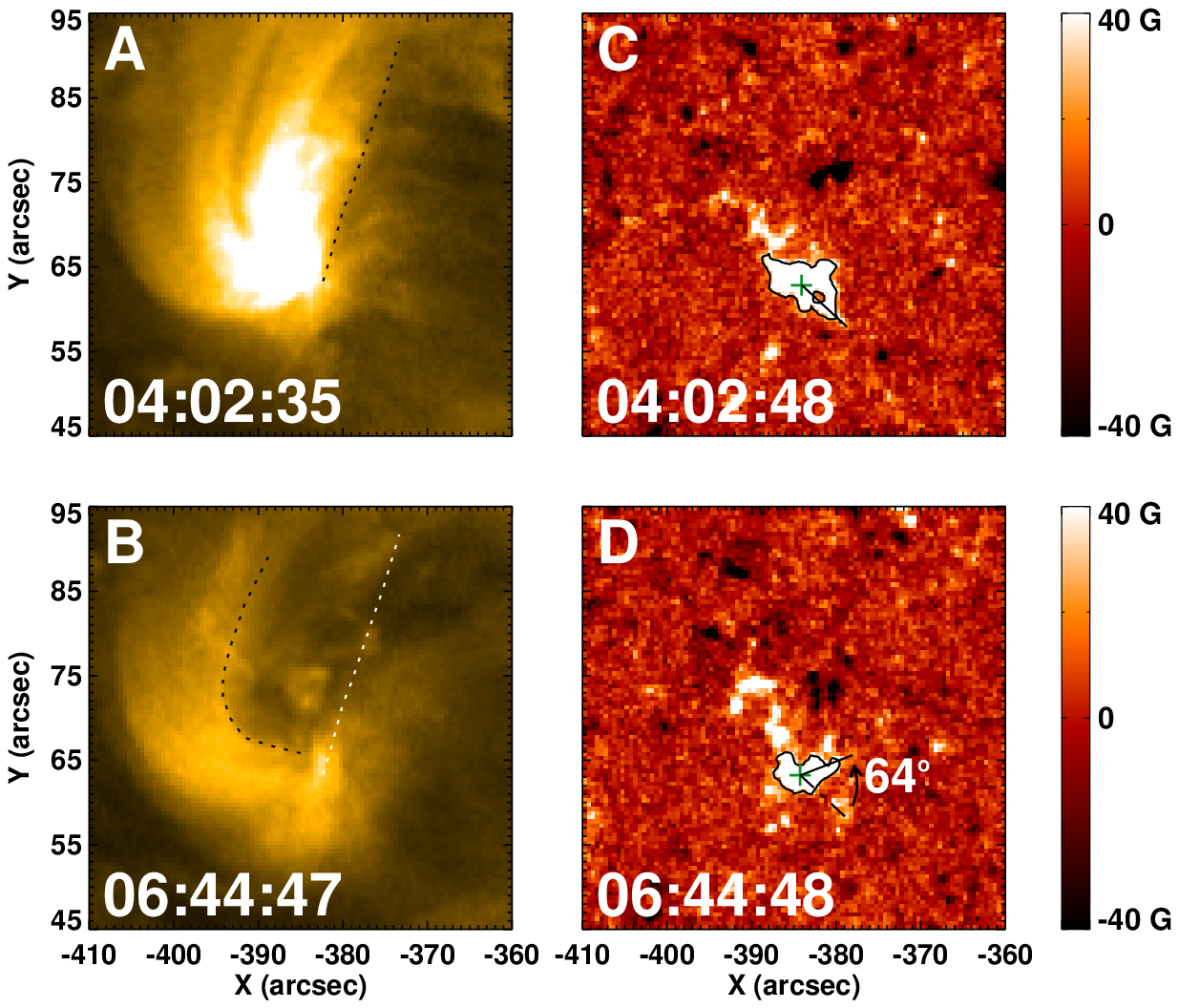} \caption{(A) and (B): Two AIA 171
{\AA} images showing a cyclone on July 2, 2010 (see the accompanying
movie1). The two black dashed curves outline one of the cyclone
boundary at the two given times, and the white dashed curve in (B)
is a duplicate of the black curve in (A). (C) and (D): Two
corresponding HMI LOS magnetograms. The lines on the magnetograms
denote the longer axis directions of the white patch which
represents a positive RNF rooted in by the cyclone. In panels (C)
and (D), the green plus symbols indicate the magnetic centroids, and
the black curves are the contours of the positive magnetic elements
at a level of 50 G. From 04:02:48 UT to 06:44:48 UT, the RNF rotated
counter-clockwise 64$^{\circ}$. For all images the North is upwards.
\label{fig1}}
\end{figure}

\begin{figure}
\vspace{-0.5cm} \epsscale{0.7} \plotone{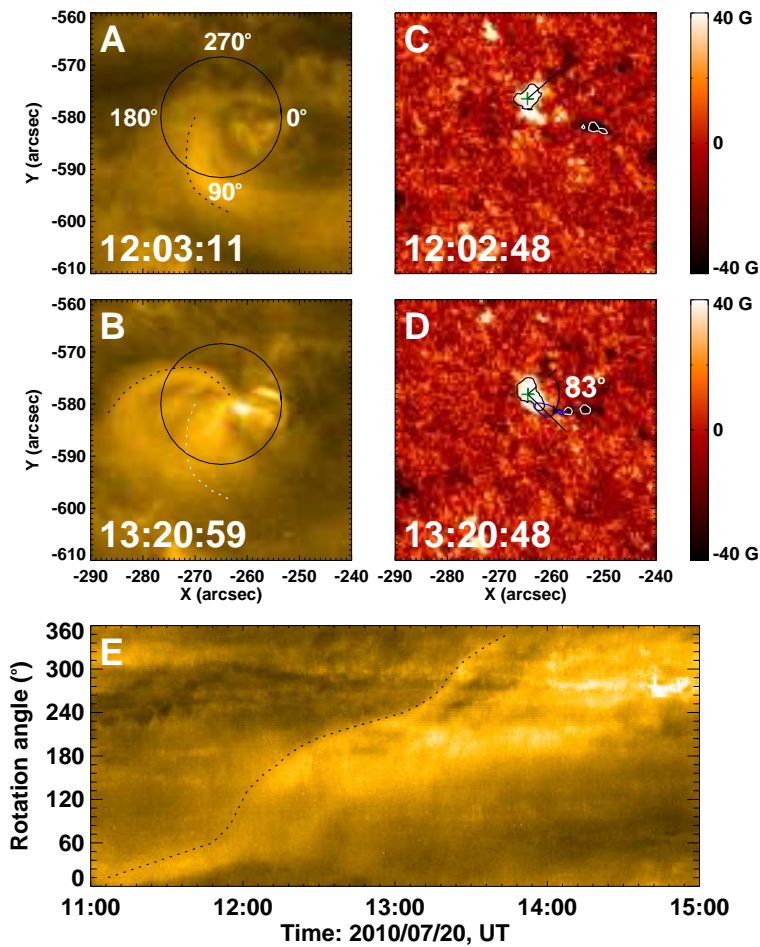} \vspace{-1.5cm}
\caption{(A) and (B): Two AIA 171 {\AA} images showing a cyclone on
July 20, 2010 (see the accompanying movie2). The two circles
represent the position at which a time-slice map is made, and the
curves represent the boundary same as that in Figures. 1(A) and
1(B). (C) and (D): Two corresponding HMI magnetograms. The lines
represent the directions similar to those in Figures. 1(C) and 1(D).
In panels (C) and (D), the green plus symbols indicate the magnetic
centroids, and the black and white curves are the contours of the
positive (50 G) and negative (-50 G) magnetic elements,
respectively. The blue curve in panel (D) is the contour of EUV
brightening in panel (B). From 12:02:48 UT to 13:20:48 UT, the RNF
rotated clockwise 83$^{\circ}$. (E): A time-slice map taken from the
AIA 171 {\AA} images. The X-axis represents the time, running from
2010 July 20, 11:00 UT to 15:00 UT. The Y-axis denotes the angle at
the center of the circle shown in panel (A), subtended by the arcs
starting at the west (right) and measured in the clockwise
direction. The dashed curve outlines the intersection of the leading
boundary of the cyclone. \label{fig2}}
\end{figure}

\begin{figure}
\epsscale{1.0} \plotone{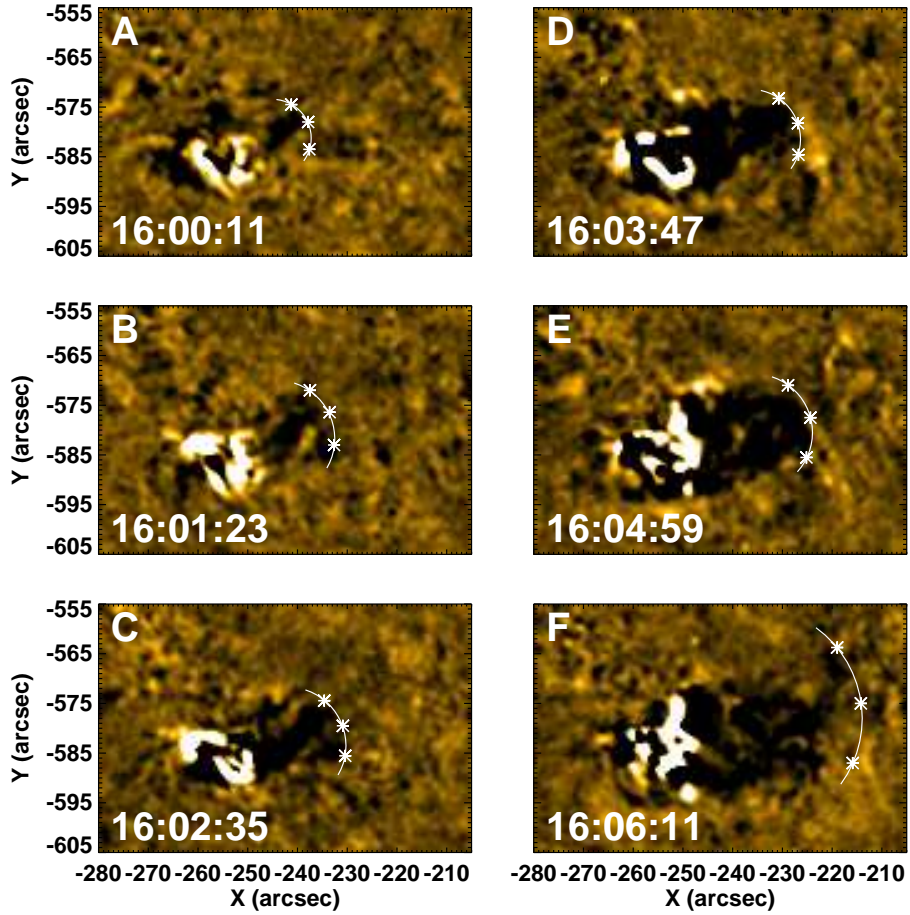} \caption{The time sequence of AIA
171 {\AA} running difference images showing a microflare (white
patches) and an EUV wave which follows the cyclone on July 20, 2010.
Three stars on each map denote the locations of the wave front, and
the propagating speed of this wave (Figure 4(B)) is determined from
these locations. \label{fig3}}
\end{figure}

\begin{figure}
\epsscale{0.8} \plotone{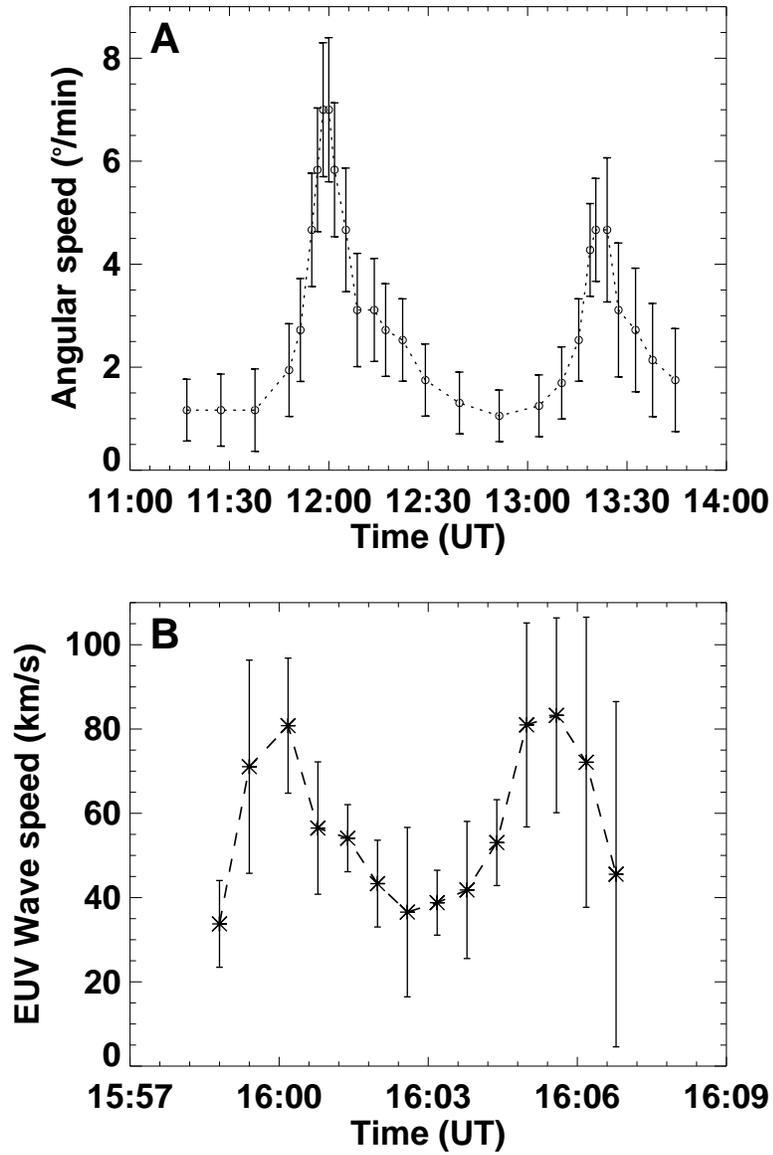}
\caption{(A): Angular speed of the cyclone on July 20, 2010. (B): The
propagating speed of the EUV wave displayed in Figure 3. \label{fig4}}
\end{figure}

\begin{figure}
\epsscale{1.0} \plotone{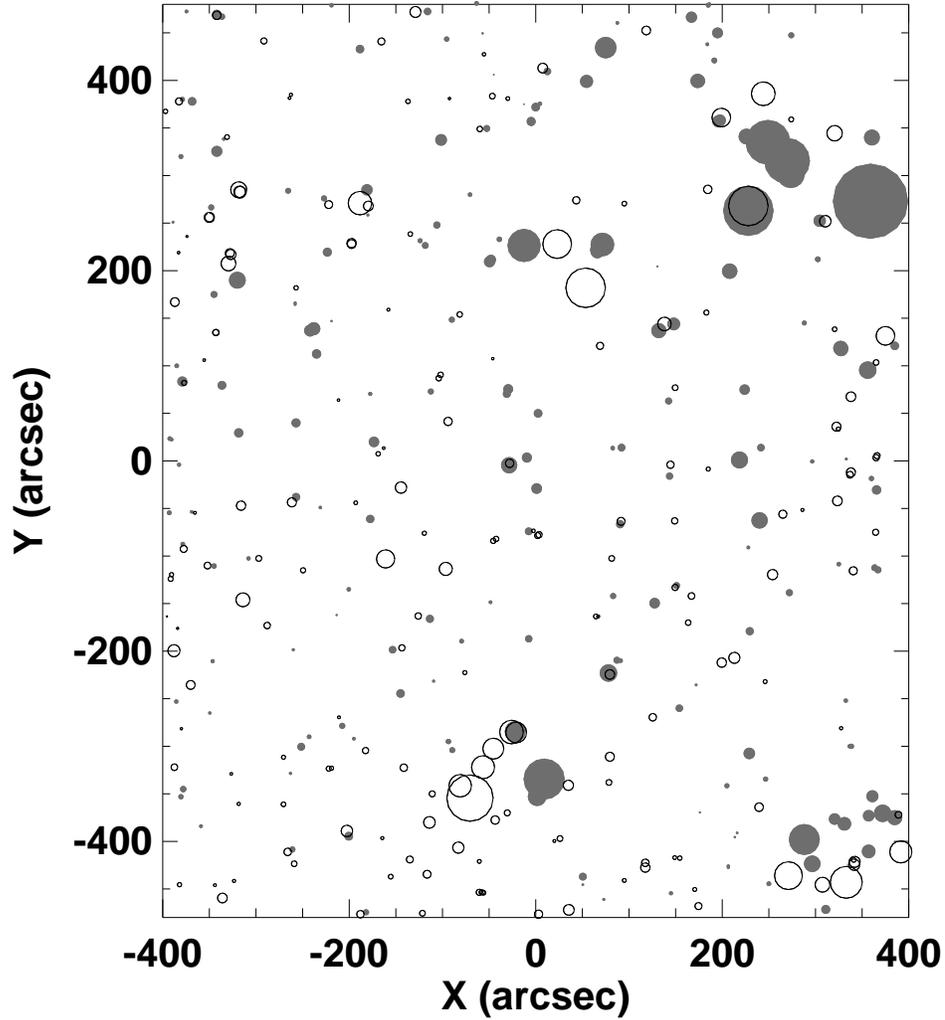}
\vspace{-2.5cm}
\caption{A plot showing the distribution of the RNFs (also see the
accompanying movie3 for an example) in the solar
disk in the FOV of 800$\times$980 square arcsec on July 8, 2010.
The size of circles represent that of the RNF flux, the largest size
5.85$\times$10$^{19}$ Mx, and the smallest 1.24$\times$10$^{17}$ Mx.
Open circles indicate that the RNFs rotate clockwise, and filled
circles, counter-clockwise. \label{fig5}}
\end{figure}

\clearpage

\end{document}